\documentclass[sigconf]{acmart}

\copyrightyear{2025}
\acmYear{2025}
\setcopyright{cc}
\setcctype{by}
\acmConference[SIGIR '25]{Proceedings of the 48th International ACM SIGIR Conference on Research and Development in Information Retrieval}{July 13--18, 2025}{Padua, Italy}
\acmBooktitle{Proceedings of the 48th International ACM SIGIR Conference on Research and Development in Information Retrieval (SIGIR '25), July 13--18, 2025, Padua, Italy}\acmDOI{10.1145/3726302.3730194}
\acmISBN{979-8-4007-1592-1/2025/07}

\AtBeginDocument{%
  }

\DeclareMathAlphabet{\pazocal}{OMS}{zplm}{m}{n}
\DeclareMathAlphabet{\pazobfcal}{OMS}{cmsy}{b}{n}

\usepackage{adjustbox}
\usepackage{multirow}
\usepackage{colortbl}
\usepackage{enumitem} 

\usepackage{subcaption}
\usepackage{enumitem}
\usepackage{hyperref}
\usepackage{array}

\newcommand{\para}[1]{\paragraph{\textnormal{\textbf{#1}}}} 
\renewcommand{\vec}[1]{\mathbf{#1}}

\usepackage{siunitx}
\newcommand{\uls}{\begin{itemize}[leftmargin=*]}
\newcommand{\ule}{\end{itemize}}
\newcommand{\ols}{\begin{enumerate}[leftmargin=*]}
\newcommand{\ole}{\end{enumerate}}
\newcommand{\li}{\item}
\begin{document}

\title{Exploring the Role of Diversity in Example Selection for In-Context Learning}

\author{Janak Kapuriya}
\authornote{Corresponding authors.}
\affiliation{%
\department{Data Science Institute}
\institution{University of Galway}
\city{Galway}
\country{Ireland}
}
\email{janakkumar.kapuriya@insight-centre.org}

\author{Manit Kaushik}
\authornotemark[1]
\affiliation{%
\department{Computer Science and Engineering}
\institution{IIIT Delhi}
\city{New Delhi}
\country{India}
}
\email{manit22277@iiitd.ac.in}

\author{Debasis Ganguly}
\affiliation{%
\department{School of Computing}
\institution{University of Glasgow}
\city{Glasgow}
\country{United Kingdom}
}
\email{Debasis.Ganguly@glasgow.ac.uk}

\author{Sumit Bhatia}
\affiliation{%
\department{Media and Data Science Research Lab}
\institution{Adobe Systems}
\city{Noida}
\country{India}
}
\email{sumit.bhatia@adobe.com}

\begin{abstract}

In-Context Learning (ICL) has gained prominence due to its ability to perform tasks without requiring extensive training data and its robustness to noisy labels. A typical ICL workflow involves selecting localized examples relevant to a given input using sparse or dense embedding-based similarity functions. However, relying solely on similarity-based selection may introduce topical biases in the retrieved contexts, potentially leading to suboptimal downstream performance.
We posit that reranking the retrieved context to enhance topical diversity can improve downstream task performance. To achieve this, we leverage maximum marginal relevance (MMR) which balances topical similarity with inter-example diversity. Our experimental results demonstrate that diversifying the selected examples leads to consistent improvements in downstream performance across various context sizes and similarity functions. The implementation of our approach is made available at \url{https://github.com/janak11111/Diverse-ICL}.

\end{abstract}



\maketitle

\section{Introduction}

Recent advances in large language models (LLMs) -- such as Mistral \cite{jiang2023mistral}, LLaMA \citep{dubey2024llama, touvron2023llama}, and various GPT variants \cite{brown2020language, achiam2023gpt}—have significantly accelerated progress in natural language processing. The increasing scale of these models has also given rise to emergent capabilities \citep{brown2020language}, notably the ability to perform tasks using only a small number of labeled examples provided in the input context, commonly known as in-context learning (ICL). ICL is conceptually similar to non-parametric learning, where instead of learning parameters during the training phase, inferences are made on unseen instances by leveraging the information from similar instances of a training set.
The performance of ICL mainly relies on the choice of localized examples relevant to a given query, especially in retrieval tasks \citep{sinhababu2024few}. A common approach involves selecting localized examples based on their similarity to the test instance using sparse or dense similarity based functions. However, this strategy often introduces topical bias by selecting examples that are also similar to each other, limiting the diversity of knowledge necessary for accurate predictions \citep{santos2010exploiting, maxwell2019impact}.

The effectiveness of ICL can potentially be influenced not only by the number of examples \cite{chandra2024one} but also their diversity. With examples related to a diverse set of topics, there is a high likelihood as per the cluster hypothesis \cite{DBLP:journals/ipm/RoyGMJ19} that some of them should turn out to be of positive utility, i.e., lead to gains in downstream performance relative to zero-shot results \cite{DBLP:conf/ecir/TianGM25}.
In contrast, a homogeneous set of examples with high inter-document similarity may lead to a ``hit or miss'' scenario, with the risk that none of them turn out to be beneficial for contextual generation.
To obtain diverse examples, we propose to rerank a top-retrieved set of candidate examples with maximal marginal relevance (MMR)  \citep{carbonell1998use}, which is a greedy algorithm that seeks to maximize the similarity of the selected examples with the input, while simultaneously minimizing the similarity between the selected examples.

\begin{figure*}[h]
  \centering
  \includegraphics[width=.87\textwidth]{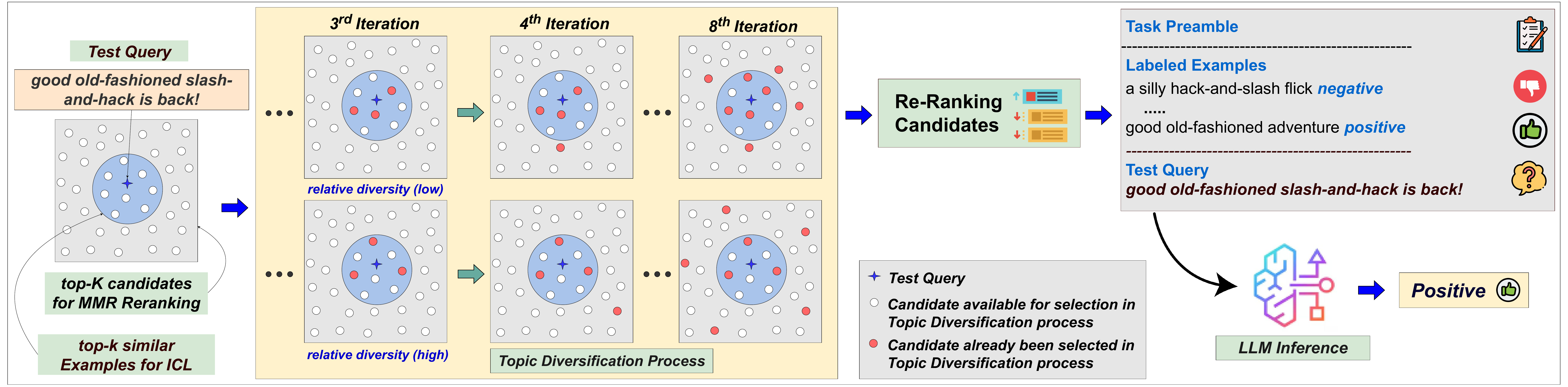}
    \caption{A schematic representation of Diversity-based In-Context Learning (DICL). The figure illustrates intermediate steps in the topic diversification process for a sample input (a sample movie review from the SST2 dataset). In this example, a total of $k=8$ candidates need to be selected. We schematically illustrate the selection mechanism employed during three different steps of the algorithm
    with two different diversity preference settings for MMR. 
    The bottom row of steps corresponds to the case when the relative importance of diversity is higher than the train-test similarity. It can be seen that this leads to a large number of examples being selected from the set $\mathcal{N}_K - \mathcal{N}_k$, i.e., from the larger pool of of size $K=nk$ of candidate examples that do not belong to the interior of the local $k$-neighborhood (the circle) and are relatively farther away from the test query. 
    }
\label{fig:main_diagram}
\end{figure*}

\section{Related Work} 
\label{sec:RelWork}

Recent advances in in-context learning (ICL) primarily aim to enhance downstream performance. For instance, \citet{rubin2022learningretrievepromptsincontext} propose a dense retrieval model for selecting relevant examples, while \citet{Parry_2024} and \citet{chandra2024one} explore dynamic adjustments using unsupervised rank-cutoff and variable example strategies. Additionally, \citet{ye2023compositional} and \citet{zhang2022active} improve candidate diversity through Determinantal Point Processes and Q-learning. However, these approaches require model training before ICL, increasing computational overhead.

Several studies emphasize the importance of diversity in ICL. \citet{Parry_2024} compares ICL with faceted IR, whereas \citet{li2023finding} proposes to filter candidates based on informativeness followed by a stochastic beam search to reduce input order sensitivity.
\citet{zhang2022automatic} proposes a clustering-based method, selecting examples from each cluster to induce diversity among the selected examples. Similarly, \citet{levy2023diversedemonstrationsimproveincontext} highlights that selecting diverse examples improves semantic parsing by encompassing all the structured patterns required for parsing. Furthermore, \citet{pickett2024better} proposed an information gain strategy to minimize redundancy in question-answering with few-shot prompts, which implicitly functions as a diversity-based example selection method. Unlike these approaches, which often require auxiliary training with diversity-focused objectives, our method directly applies the MMR approach to rerank a top-retrieved set of candidate examples. This approach inherently promotes topical diversity without additional training, offering a more efficient and scalable solution.

\section{Diversity-based In-Context Learning (DICL)}
\label{sec:MMR_methodology}
We propose a diversity-based ICL (DICL) method that balances the relevance of examples (similarity with input instance) and their diversity. We now describe the key steps in our proposed approach.

\para{Standard ICL}
In ICL, an LLM is used to predict the posterior distribution over target classes, which is a function of a) the embedding $\vec{x}$ of a test instance, b) the (frozen) decoder parameters $\phi$ of a pre-trained LLM, c) the prompt instruction, and, more importantly, d) a set -- $\mathcal{N}_k(\vec{x}; \theta)$ -- of $k$ labeled examples from a training set.
Formally, $P(y|\vec{x};\phi,\theta) = f(\vec{x}, \mathcal{N}_k(\vec{x}; \theta);\phi)$, where $\theta$ denotes a similarity 
function used to induce an ordering on each training set instance prior to selecting the top candidates. This can either be a lexical similarity function that relies on exact term matches, defined on bag-of-words representation of text, e.g., tf-idf, or a semantic similarity function, as defined over dense vector representations of text, which is not restricted to strict word matches.

\para{Selecting candidate examples}
Since the proposed diversity-based approach involves reranking a set of local neighborhood of most similar examples - $\mathcal{N}(\vec{x}; \theta)$, we first start with retrieving a pool of candidate examples that is larger in size than the number of examples eventually used in ICL inference. Specifically, for $k$-shot inference, the idea is to first obtain a candidate set of $K = nk$ examples, where $n$ is a small integer.
We denote this larger neighborhood with the notation $\mathcal{N}_K(\vec{x}; \theta) \supset \mathcal{N}_k(\vec{x}; \theta)$, e.g., Figure \ref{fig:main_diagram} depicts this larger neighborhood of candidates with a square, whereas the smaller (interior) one is shown with a circle. 

Specifically, we set $n=3$ in all our experiments as this value balances between ensuring a sufficiently large and diverse pool while avoiding the inclusion of non-relevant examples.

\para{MMR for DICL}
To re-rank the set of size $K=nk$ candidates with an objective to balance train-test similarity and diversity, we apply the MMR algorithm \cite{10.1145/290941.291025}. Specifically, MMR is an iterative algorithm which maintains an incrementally growing list of selected instances, and the score is a function of
a) the step count $i$ ($i < k$ - the number of instances to be selected), b) the set of instances selected till the $(i-1)^{\text{th}}$ step, denoted by $S_{i-1}=\{ \vec{s}_1,\ldots, \vec{s}_{i-1}\}$, and c) $\alpha \in [0,1]$, a hyper-parameter that controls the relative importance of relevance with respect to diversity. 
Formally, 
\begin{equation}
\theta_{\text{MMR}}(\vec{x}, \vec{u}, S_{i-1}; \alpha)
=
\alpha \theta(\vec{x}, \vec{u}) -
(1 - \alpha) \max_{\vec{s} \in S_{i-1}} \theta(\vec{u}, \vec{s}),\,\,
\forall \vec{u} \in \mathcal{N}_K(\vec{x}),
\label{eq:mmr}
\end{equation}
where $\theta$ is a similarity function (lexical or semantic) between instances. Executing Equation \ref{eq:mmr} for $k$ steps yields the desired reranked list of $k$ examples from a larger candidate set of size $K=nk$. The working principle of Equation \ref{eq:mmr} is illustrated in Figure \ref{fig:main_diagram}.

\definecolor{darkgreen}{RGB}{0,128,0}
\begin{table*}[t]
\centering
\caption{DICL performance reported
for 4 datasets across 3 LLMs.
The symbol \textcolor{darkgreen}{↑X} (\textcolor{red}{↓X}) denotes percentage increase (decrease) in DICL performance as compared to standard ICL. The superscripts $a$ and $b$ respectively indicate statistically significant (paired t-test with $p=0.05$) improvements over the 0-shot and standard ICL baselines.
}
\small
\begin{tabular}{@{}ll*{3}{cclc}@{}}
\toprule
Dataset & Method
  & \multicolumn{4}{c}{Phi2-2.7B}
  & \multicolumn{4}{c}{Mistral-7B}
  & \multicolumn{4}{c}{LLaMA3-8B} \\
\cmidrule(r){3-6} \cmidrule(r){7-10} \cmidrule(r){11-14}
& 
  & $K$ & $\alpha$ & \multicolumn{1}{c}{F1} & \multicolumn{1}{c}{$\Delta(\%)$}
  & $K$ & $\alpha$ & \multicolumn{1}{c}{F1} & \multicolumn{1}{c}{$\Delta(\%)$}
  & $K$ & $\alpha$ & \multicolumn{1}{c}{F1} & \multicolumn{1}{c}{$\Delta(\%)$} \\
\midrule
\multirow{6}{*}{RTE}
 & 0-Shot      & 0 & N/A & 0.4870 & N/A & 0 & N/A & 0.5211 & N/A & 0 & N/A & 0.5860 & N/A \\ %
 & TFIDF     & 1 & N/A & \textbf{0.5503} & N/A & 9 & N/A & \textbf{0.7515}\textsuperscript{$a$} & N/A & 5 & N/A & 0.7074\textsuperscript{$a$} & N/A \\
 & \cellcolor{gray!25}TFIDF-MMR & \cellcolor{gray!25}1 & \cellcolor{gray!25}0.9 & \cellcolor{gray!25}\textbf{0.5649} & \cellcolor{gray!25} \textcolor{darkgreen}{$\uparrow$ 2.7} & \cellcolor{gray!25}7 & \cellcolor{gray!25}0.8 & \cellcolor{gray!25}0.7484\textsuperscript{$a$} & \cellcolor{gray!25} \textcolor{red} {$\downarrow$ 0.4} & \cellcolor{gray!25}9 & \cellcolor{gray!25}0.9 & \cellcolor{gray!25}\textbf{0.7173}\textsuperscript{$a$} & \cellcolor{gray!25} \textcolor{darkgreen}{$\uparrow$ 1.4} \\ 
 & SBERT      & 1 & N/A & 0.5464 & N/A & 5 & N/A & 0.7161\textsuperscript{$a$} & N/A & 10 & N/A & 0.6919\textsuperscript{$a$} & N/A \\
 & \cellcolor{gray!25}SBERT-MMR  & \cellcolor{gray!25}1 & \cellcolor{gray!25}0.9 & \cellcolor{gray!25}0.5464 & \cellcolor{gray!25} \textcolor{gray}{ 0.0}  & \cellcolor{gray!25}10 & \cellcolor{gray!25}0.5 & \cellcolor{gray!25}\textbf{0.7320}\textsuperscript{$a$} & \cellcolor{gray!25} \textcolor{darkgreen}{$\uparrow$ 2.2}  & \cellcolor{gray!25}9 & \cellcolor{gray!25}0.7 & \cellcolor{gray!25}\textbf{0.7133}\textsuperscript{$a$} & \cellcolor{gray!25} \textcolor{darkgreen}{$\uparrow$ 3.1} \\
\midrule
\multirow{6}{*}{COLA}
 & 0-Shot      & 0 & N/A & 0.3261 & N/A & 0 & N/A & 0.5566 & N/A& 0 & N/A & 0.5734 & N/A\\ %
 & TFIDF     & 1 & N/A & 0.3244\textsuperscript{$a$} & N/A & 7 & N/A & \textbf{0.7729}\textsuperscript{$a$} & N/A & 10 & N/A & \textbf{0.6841}\textsuperscript{$a$} & N/A\\
 & \cellcolor{gray!25}TFIDF-MMR & \cellcolor{gray!25}10 & \cellcolor{gray!25}0.5 & \cellcolor{gray!25}\textbf{0.3531}\textsuperscript{$a, b$} & \cellcolor{gray!25} \textcolor{darkgreen}{$\uparrow$ 8.8} & \cellcolor{gray!25}7 & \cellcolor{gray!25}0.7 & \cellcolor{gray!25}0.7670\textsuperscript{$a$} & \cellcolor{gray!25} \textcolor{red}{$\downarrow$ 0.8} & \cellcolor{gray!25}10 & \cellcolor{gray!25}0.8 & \cellcolor{gray!25}0.6751\textsuperscript{$a$} & \cellcolor{gray!25} \textcolor{red}{$\downarrow$ 1.3} \\ %
 & SBERT      & 1 & N/A & 0.3278\textsuperscript{$a$} & N/A & 7 & N/A & 0.7643\textsuperscript{$a$} & N/A & 9 & N/A & 0.6909\textsuperscript{$a$} & N/A \\
 & \cellcolor{gray!25}SBERT-MMR  & \cellcolor{gray!25}10 & \cellcolor{gray!25}0.5 & \cellcolor{gray!25}\textbf{0.3598}\textsuperscript{$a, b$} & \cellcolor{gray!25} \textcolor{darkgreen}{$\uparrow$ 9.7}  & \cellcolor{gray!25}7 & \cellcolor{gray!25}0.8 & \cellcolor{gray!25}\textbf{0.7851}\textsuperscript{$a, b$} & \cellcolor{gray!25} \textcolor{darkgreen}{$\uparrow$ 2.7}  & \cellcolor{gray!25}7 & \cellcolor{gray!25}0.5 & \cellcolor{gray!25}\textbf{0.6925}\textsuperscript{$a$} & \cellcolor{gray!25} \textcolor{darkgreen}{$\uparrow$ 0.2}\\
\midrule
\multirow{6}{*}{SST2}
 & 0-Shot      & 0 & N/A & 0.8587 & N/A & 0 & N/A & 0.8904 & N/A & 0 & N/A & 0.7509 & N/A\\ %
 & TFIDF     & 7 & N/A & 0.9151\textsuperscript{$a$} & N/A & 9 & N/A & 0.9426\textsuperscript{$a$} & N/A & 9 & N/A & 0.8842\textsuperscript{$a$} & N/A \\
 & \cellcolor{gray!25}TFIDF-MMR & \cellcolor{gray!25}7 & \cellcolor{gray!25}0.9 & \cellcolor{gray!25}0.9151\textsuperscript{$a$} & \cellcolor{gray!25} \textcolor{gray}{ 0.0} & \cellcolor{gray!25}9 & \cellcolor{gray!25}0.4 & \cellcolor{gray!25}\textbf{0.9438}\textsuperscript{$b$} & \cellcolor{gray!25} \textcolor{darkgreen}{$\uparrow$ 0.1} & \cellcolor{gray!25}3 & \cellcolor{gray!25}0.5 & \cellcolor{gray!25}\textbf{0.8853}\textsuperscript{$a$} & \cellcolor{gray!25} \textcolor{darkgreen}{$\uparrow$ 0.1} \\ %
 & SBERT      & 9 & N/A & \textbf{0.9255}\textsuperscript{$a$} & N/A & 9 & N/A & \textbf{0.9484}\textsuperscript{$a$} & N/A & 7 & N/A & 0.9220\textsuperscript{$a$} & N/A \\
 & \cellcolor{gray!25} SBERT-MMR  & \cellcolor{gray!25}9 & \cellcolor{gray!25}0.9 & \cellcolor{gray!25}0.9220\textsuperscript{$a$} & \cellcolor{gray!25} \textcolor{red}{$\downarrow$ 0.4} & \cellcolor{gray!25}9 & \cellcolor{gray!25}0.4 & \cellcolor{gray!25}0.9472\textsuperscript{$a$} & \cellcolor{gray!25} \textcolor{red}{$\downarrow$ 0.1} & \cellcolor{gray!25}9 & \cellcolor{gray!25}0.4 & \cellcolor{gray!25}\textbf{0.9335}\textsuperscript{$a$} & \cellcolor{gray!25} \textcolor{darkgreen}{$\uparrow$ 1.2} \\
\midrule
\multirow{6}{*}{TREC} 
 & 0-Shot      & 0 & N/A & 0.4967 & N/A & 0 & N/A & 0.1955 & N/A  & 0 & N/A & 0.5935 & N/A \\ %
 & TFIDF     & 7 & N/A & \textbf{0.8446}\textsuperscript{$a$} & N/A & 7 & N/A & 0.8780\textsuperscript{$a$} & N/A & 7 & N/A & 0.7599\textsuperscript{$a$} & N/A \\
 & \cellcolor{gray!25}TFIDF-MMR & \cellcolor{gray!25}10 & \cellcolor{gray!25}0.8 & \cellcolor{gray!25}0.8222\textsuperscript{$a$} & \cellcolor{gray!25} \textcolor{red}{$\downarrow$ 2.6}  & \cellcolor{gray!25}7 & \cellcolor{gray!25}0.9 & \cellcolor{gray!25}\textbf{0.9066}\textsuperscript{$a, b$} & \cellcolor{gray!25} \textcolor{darkgreen}{$\uparrow$ 3.2}  & \cellcolor{gray!25}10 & \cellcolor{gray!25}0.9 & \cellcolor{gray!25}\textbf{0.8268}\textsuperscript{$a, b$} & \cellcolor{gray!25} \textcolor{darkgreen}{$\uparrow$ 8.8} \\ %
 & SBERT      & 5 & N/A & 0.8070\textsuperscript{$a$} & N/A & 9 & N/A & 0.8418\textsuperscript{$a$} & N/A & 10 & N/A & \textbf{0.7606}\textsuperscript{$a$} & N/A \\
 & \cellcolor{gray!25}SBERT-MMR  & \cellcolor{gray!25}10 & \cellcolor{gray!25}0.8 & \cellcolor{gray!25}\textbf{0.8304}\textsuperscript{$a$} & \cellcolor{gray!25} \textcolor{darkgreen}{$\uparrow$ 2.9} & \cellcolor{gray!25}10 & \cellcolor{gray!25}0.6 & \cellcolor{gray!25}\textbf{0.8459}\textsuperscript{$a$} & \cellcolor{gray!25} \textcolor{darkgreen}{$\uparrow$ 0.5} & \cellcolor{gray!25}9 & \cellcolor{gray!25}0.9 & \cellcolor{gray!25}0.7283\textsuperscript{$a$} & \cellcolor{gray!25} \textcolor{red}{$\downarrow$ 4.2} \\
\bottomrule
\label{tab:ICL_results_table}
\end{tabular}
\end{table*}

\section{Experiment Setup}
\para{Research Questions}
We conducted experiments to answer the following research questions.
\uls
\li \textbf{RQ1}: Does the Diversity-based ICL (DICL) method lead to better downstream performance than standard ICL methods involving lexical or semantic similarities?
\li \textbf{RQ2}: How sensitive is DICL to its hyper-parameters, i.e., $\alpha$ - the relevance:diversity preference, and $k$ - the number of labeled examples?
\li \textbf{RQ3}: Does DICL exhibit consistent improvements across LLMs of different families and sizes?
\ule

\para{Datasets} For our experiments, we employed four datasets: RTE, COLA, SST2, and TREC with the former three
(corresponding to the tasks of textual entailment, grammatical acceptability, and sentiment classification) being a part of the GLUE benchmark \citep{wang2018glue}. In contrast,
the TREC dataset corresponds to the task of open-domain question classification \citep{ma2023fairness}.

\para{Methods Investigated}
We compare the following methods.

\uls
\li \textbf{Zero-Shot}: This method does not leverage examples for predictions, and relies on the internal knowledge representation capabilities of an LLM.

\li \textbf{TFIDF}: Uses tf-idf as the similarity function $\theta$ to construct the neighborhood of $k$ most similar examples from the training set.
\li \textbf{SBERT:} Similar to TF-IDF, the only difference being that as similarity function it uses cosine similarities computed over dense vectors (Sentence-BERT embeddings \citep{reimers2019sentence})
\li \textbf{TFIDF-MMR}: The DICL variant of standard ICL where the initial set of candidate examples are obtained with the tf-idf similarity, subsequently reranked by Equation \ref{eq:mmr}.
\li \textbf{SBERT-MMR}: Similar to TF-IDF MMR, the difference being dense vectors are used for similarity computation during the initial step and MMR reranking.
\ule

\begin{figure*}[t]
\captionsetup[subfigure]{justification=centering} 
  \begin{subfigure}[b]{0.24\textwidth}
    \includegraphics[width=\textwidth]{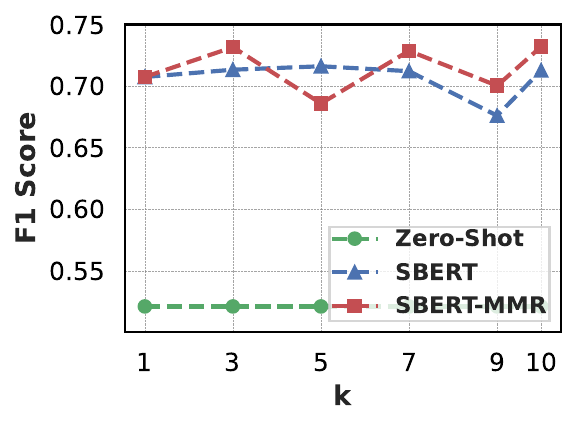}
    \caption{Mistral-RTE-SBERT}
    \label{rte_mistral_sbert}
  \end{subfigure}%
  \begin{subfigure}[b]{0.24\textwidth}
    \includegraphics[width=\textwidth]{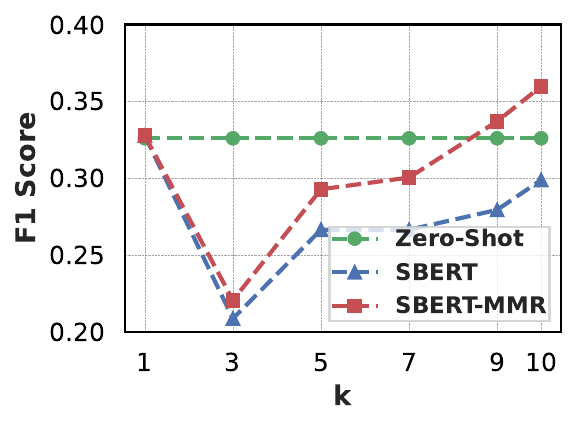}
    \caption{Phi2-COLA-SBERT}
    \label{cola_phi_sbert}
  \end{subfigure}%
  \begin{subfigure}[b]{0.24\textwidth}
    \includegraphics[width=\textwidth]{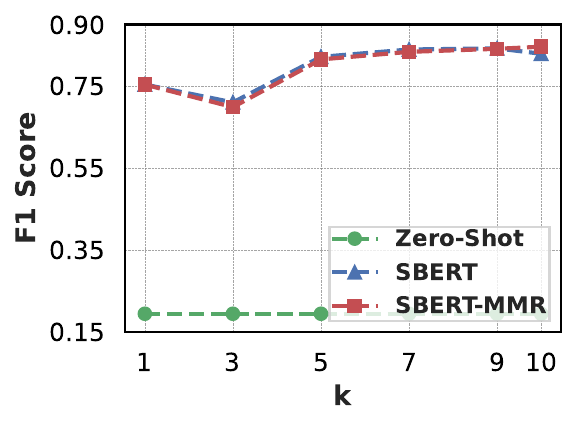}
    \caption{Mistral-TREC-SBERT}
    \label{trec_mistral_sbert}
  \end{subfigure}%
  \begin{subfigure}[b]{0.24\textwidth}
    \includegraphics[width=\textwidth]{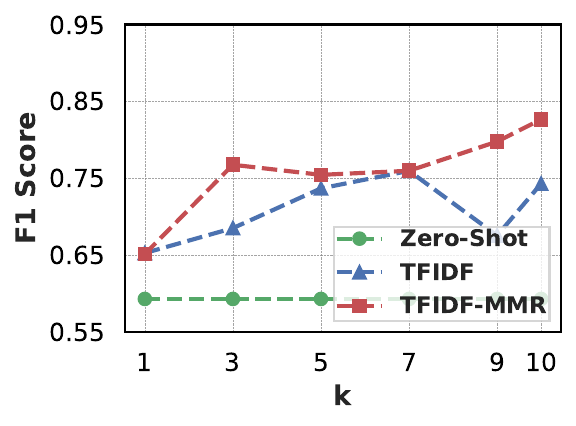}
    \caption{LLaMA-TREC-TFIDF}
    \label{trec_llama_tfidf}
  \end{subfigure}
  \begin{subfigure}[b]{0.24\textwidth}
    \includegraphics[width=\textwidth]{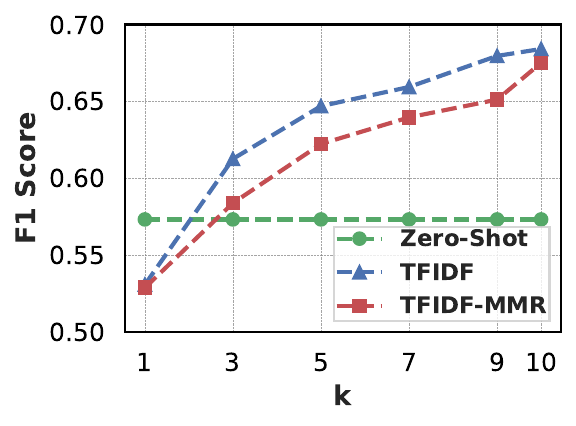}
    \caption{LLaMA-COLA-TFIDF}
    \label{cola_llama_tfidf}
  \end{subfigure}
  \begin{subfigure}[b]{0.24\textwidth}
    \includegraphics[width=\textwidth]{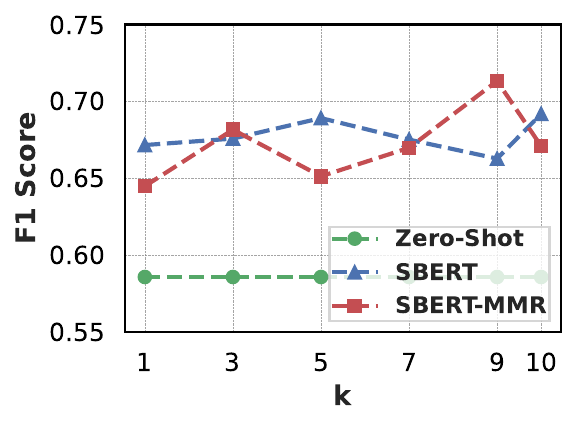}
    \caption{LLaMA-RTE-SBERT}
    \label{rte_llama_sbert}
  \end{subfigure}
  \begin{subfigure}[b]{0.24\textwidth}
    \includegraphics[width=\textwidth]{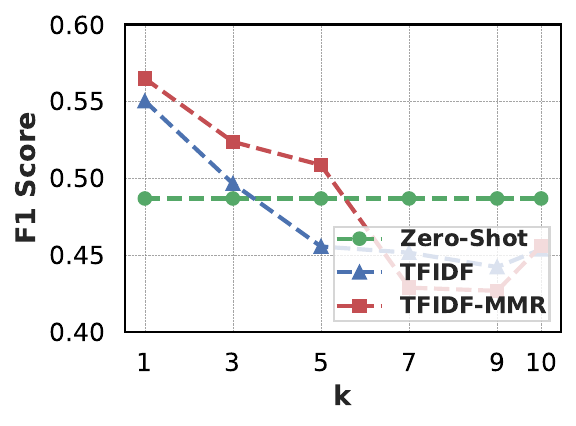}
    \caption{Phi2-RTE-TFIDF}
    \label{rte_phi2_tfidf}
  \end{subfigure}
  \begin{subfigure}[b]{0.24\textwidth}
    \includegraphics[width=\textwidth]{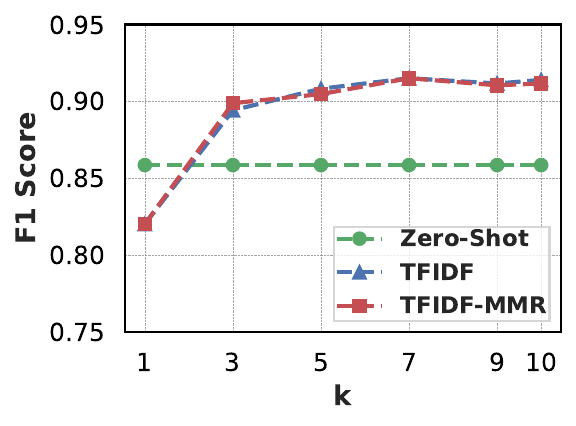}
    \caption{Phi2-SST2-TFIDF}
    \label{sst_phi2_tfidf}
  \end{subfigure}
  \caption{Effect of varying the number of examples in ICL and DICL across 4 datasets and 3 LLMs.
  }
  \label{fig:CapSFT}
  \end{figure*}

\para{Parameters and Configurations}
We employ LLMs from three different model families: \texttt{Mistral-7B-v0.3} \citep{jiang2023mistral}, \texttt{LLaMA-3.1-8B} \citep{dubey2024llama} and 
 \texttt{Phi-2-2.7B} \citep{li2023textbooks}. These models differ in their architecture, size and pretraining corpus. All models were used in their quantized 4-bit versions \citep{dettmers2024qlora}.
All hyper-parameters ($\alpha$ and $k$ in DICL, vs. only $k$ in ICL) were tuned by conducting a grid search on the validation splits of each respective dataset.
In particular, for grid search we considered $\alpha \in \{0, 0.1,\ldots, 0.9\}$ ($\alpha=1$ degenerates to standard ICL), and $k \in \{1, 3, 5, 7,     9, 10 \}$.
These optimal settings for each method were then used to evaluate and report the results on the test splits.

\section{Results and Discussions}

Table \ref{tab:ICL_results_table} reports our main findings across a range of different experiment configurations as defined by the choice of the LLMs, similarity functions, and whether or not diversity was used in ICL.
We now present the observations aligned with our research questions.

\para{RQ1: DICL Downstream Performance}
For RQ-1, it can be observed that the MMR-based approach consistently outperforms or matches the baseline similarity-based retrieval in 17 out of 24 such settings, indicating that incorporating diversity enhances or at least preserves performance in approximately 70\% of the cases. 

\para{RQ2: DICL Hyperparameter Sensitivity}  
On the RTE dataset, nearly all configurations either improve or maintain performance when enhanced with MMR-based diversity. As shown in Figures \hyperref[rte_mistral_sbert]{2(a)}, \hyperref[rte_llama_sbert]{2(f)} and \hyperref[rte_phi2_tfidf]{2(g)}, the inclusion of both lexical and semantic diversity helps the models better handle relational inference tasks by providing complementary contextual signals. 

For the COLA dataset, SBERT-MMR consistently outperforms TFIDF-MMR, likely due to the richer semantic representations offered by SBERT. These embeddings provide more meaningful context, which appears beneficial for tasks requiring judgments on grammatical acceptability, as shown in Figures \hyperref[cola_phi_sbert]{2(b)} and \hyperref[cola_llama_tfidf]{2(e)}. 

In sentiment classification on SST2, where sentiment cues are often strong and unambiguous, the MMR-based diversity approach offers only marginal gains. As shown in Figure \hyperref[sst_phi2_tfidf]{2(h)}, the baseline similarity-based retrieval already provides sufficient context, leaving limited room for improvement through additional diversity. 

For the TREC dataset, TFIDF-MMR, when paired with larger models such as Mistral-7B and LLaMA3-8B, yields substantial improvements over SBERT-MMR. As shown in Figures \hyperref[trec_mistral_sbert]{2(c)} and \hyperref[trec_llama_tfidf]{2(d)}, this can be attributed to the nature of open-domain question classification, where exact term overlap between questions and training examples plays a key role in identifying the correct category. In such scenarios, TFIDF's emphasis on word frequency and term specificity proves more effective than SBERT’s dense semantic representations, which may abstract away critical lexical distinctions. In contrast, the smaller Phi-2 2.7B model benefits more from SBERT-MMR, likely due to its limited capacity to resolve ambiguity without additional semantic context.

In most cases where MMR-based methods outperform their similarity-only counterparts, i) the he relevance coefficient $\alpha$ is relatively high (typically $\alpha \geq 0.5$ which means that relative importance of diversity is lower), and ii) the number of labeled examples is relatively on the higher side, e.g., $k \geq 7$. This trend suggests that diversity becomes more impactful when the prompt includes a sufficiently large example set, as smaller sets inherently limit the scope for incorporating varied information. Figures \hyperref[rte_mistral_sbert]{2(a)}, \hyperref[cola_phi_sbert]{2(b)}, \hyperref[trec_llama_tfidf]{2(d)}, and \hyperref[rte_llama_sbert]{2(f)} support this pattern, and the corresponding values of $K$ and $\alpha$ are reported in Table \ref{tab:ICL_results_table}.

\para{RQ3: DICL Sensitivity on LLM Family and Scale} For both Mistral and LLaMA, F1-scores generally improve with increasing values of $k$ before plateauing, an effect observed in both standard ICL and diversity-based ICL settings. This indicates that larger models benefit from broader context, but the marginal gains diminish beyond a certain number of examples. This can be seen both on Mistral (Figures \hyperref[rte_mistral_sbert]{2(a)} and \hyperref[trec_mistral_sbert]{2(c)}), and Llama (Figures \hyperref[trec_llama_tfidf]{2(d)}, \hyperref[cola_llama_tfidf]{2(e)}, and \hyperref[rte_llama_sbert]{2(f)}).
A similar trend is observed for the smaller class of model - Phi2 2.7B on the TREC and SST2 tasks, as shown in Figure \hyperref[sst_phi2_tfidf]{2(h)}. In contrast, on
the RTE and COLA datasets the Phi-2 model shows improvements with either a small or a large number of examples with noticeable decrease in performance in the middle range (Figures \hyperref[rte_phi2_tfidf]{2(g)} and \hyperref[cola_phi_sbert]{2(b)}).

\para{Concluding Remarks}
We explored diversity-based approaches for In-Context Learning (ICL), and our experiments revealed consistent enhancements in model generalization.
Smaller diversity preferences (\(\alpha \geq 0.5\)) and larger number of ICL examples (\(k \geq 7\)) turned out to be effective for scaling model performance, especially in larger architectures. In contrast, smaller models were more sensitive to noise, underscoring the need to carefully balance diversity and dataset size relative to model capacity.
In future, we plan to explore the role of diversity for unlabeled contexts, as encountered in retrieval augmented generation (RAG) tasks, such as question answering \cite{DBLP:conf/ecir/TianGM25,pickett2024better}.

\para{Acknowledgments}
The first author is supported by Research Ireland under Grant Number SFI/12/RC/2289\_P2 (Insight), co-funded by the European Regional Development Fund. We acknowledge the use of generative AI to improve the text quality.

\bibliographystyle{ACM-Reference-Format}
\bibliography{sample-base-updated}


\begin{thebibliography}{26}


\ifx \showCODEN    \undefined \def \showCODEN     #1{\unskip}     \fi
\ifx \showISBNx    \undefined \def \showISBNx     #1{\unskip}     \fi
\ifx \showISBNxiii \undefined \def \showISBNxiii  #1{\unskip}     \fi
\ifx \showISSN     \undefined \def \showISSN      #1{\unskip}     \fi
\ifx \showLCCN     \undefined \def \showLCCN      #1{\unskip}     \fi
\ifx \shownote     \undefined \def \shownote      #1{#1}          \fi
\ifx \showarticletitle \undefined \def \showarticletitle #1{#1}   \fi
\ifx \showURL      \undefined \def \showURL       {\relax}        \fi
\providecommand\bibfield[2]{#2}
\providecommand\bibinfo[2]{#2}
\providecommand\natexlab[1]{#1}
\providecommand\showeprint[2][]{arXiv:#2}

\bibitem[Achiam et~al\mbox{.}(2023)]%
        {achiam2023gpt}
\bibfield{author}{\bibinfo{person}{Josh Achiam}, \bibinfo{person}{Steven Adler}, \bibinfo{person}{Sandhini Agarwal}, \bibinfo{person}{Lama Ahmad}, \bibinfo{person}{Ilge Akkaya}, \bibinfo{person}{Florencia~Leoni Aleman}, \bibinfo{person}{Diogo Almeida}, \bibinfo{person}{Janko Altenschmidt}, \bibinfo{person}{Sam Altman}, \bibinfo{person}{Shyamal Anadkat}, {et~al\mbox{.}}} \bibinfo{year}{2023}\natexlab{}.
\newblock \showarticletitle{Gpt-4 technical report}.
\newblock \bibinfo{journal}{\emph{ArXiv preprint}}  \bibinfo{volume}{abs/2303.08774} (\bibinfo{year}{2023}).
\newblock
\urldef\tempurl%
\url{https://arxiv.org/abs/2303.08774}
\showURL{%
\tempurl}


\bibitem[Brown et~al\mbox{.}(2020)]%
        {brown2020language}
\bibfield{author}{\bibinfo{person}{Tom~B. Brown}, \bibinfo{person}{Benjamin Mann}, \bibinfo{person}{Nick Ryder}, \bibinfo{person}{Melanie Subbiah}, \bibinfo{person}{Jared Kaplan}, \bibinfo{person}{Prafulla Dhariwal}, \bibinfo{person}{Arvind Neelakantan}, \bibinfo{person}{Pranav Shyam}, \bibinfo{person}{Girish Sastry}, \bibinfo{person}{Amanda Askell}, \bibinfo{person}{Sandhini Agarwal}, \bibinfo{person}{Ariel Herbert{-}Voss}, \bibinfo{person}{Gretchen Krueger}, \bibinfo{person}{Tom Henighan}, \bibinfo{person}{Rewon Child}, \bibinfo{person}{Aditya Ramesh}, \bibinfo{person}{Daniel~M. Ziegler}, \bibinfo{person}{Jeffrey Wu}, \bibinfo{person}{Clemens Winter}, \bibinfo{person}{Christopher Hesse}, \bibinfo{person}{Mark Chen}, \bibinfo{person}{Eric Sigler}, \bibinfo{person}{Mateusz Litwin}, \bibinfo{person}{Scott Gray}, \bibinfo{person}{Benjamin Chess}, \bibinfo{person}{Jack Clark}, \bibinfo{person}{Christopher Berner}, \bibinfo{person}{Sam McCandlish}, \bibinfo{person}{Alec Radford}, \bibinfo{person}{Ilya Sutskever},
  {and} \bibinfo{person}{Dario Amodei}.} \bibinfo{year}{2020}\natexlab{}.
\newblock \showarticletitle{Language Models are Few-Shot Learners}. In \bibinfo{booktitle}{\emph{Advances in Neural Information Processing Systems 33: Annual Conference on Neural Information Processing Systems 2020, NeurIPS 2020, December 6-12, 2020, virtual}}, \bibfield{editor}{\bibinfo{person}{Hugo Larochelle}, \bibinfo{person}{Marc'Aurelio Ranzato}, \bibinfo{person}{Raia Hadsell}, \bibinfo{person}{Maria{-}Florina Balcan}, {and} \bibinfo{person}{Hsuan{-}Tien Lin}} (Eds.).
\newblock
\urldef\tempurl%
\url{https://proceedings.neurips.cc/paper/2020/hash/1457c0d6bfcb4967418bfb8ac142f64a-Abstract.html}
\showURL{%
\tempurl}


\bibitem[Carbonell and Goldstein(1998a)]%
        {carbonell1998use}
\bibfield{author}{\bibinfo{person}{Jaime Carbonell} {and} \bibinfo{person}{Jade Goldstein}.} \bibinfo{year}{1998}\natexlab{a}.
\newblock \showarticletitle{The use of MMR, diversity-based reranking for reordering documents and producing summaries}. In \bibinfo{booktitle}{\emph{Proceedings of the 21st annual international ACM SIGIR conference on Research and development in information retrieval}}. \bibinfo{pages}{335--336}.
\newblock


\bibitem[Carbonell and Goldstein(1998b)]%
        {10.1145/290941.291025}
\bibfield{author}{\bibinfo{person}{Jaime Carbonell} {and} \bibinfo{person}{Jade Goldstein}.} \bibinfo{year}{1998}\natexlab{b}.
\newblock \showarticletitle{The use of MMR, diversity-based reranking for reordering documents and producing summaries}. In \bibinfo{booktitle}{\emph{Proceedings of the 21st Annual International ACM SIGIR Conference on Research and Development in Information Retrieval}} (Melbourne, Australia) \emph{(\bibinfo{series}{SIGIR '98})}. \bibinfo{publisher}{Association for Computing Machinery}, \bibinfo{address}{New York, NY, USA}, \bibinfo{pages}{335–336}.
\newblock
\showISBNx{1581130155}
\href{https://doi.org/10.1145/290941.291025}{doi:\nolinkurl{10.1145/290941.291025}}


\bibitem[Chandra et~al\mbox{.}(2025)]%
        {chandra2024one}
\bibfield{author}{\bibinfo{person}{Manish Chandra}, \bibinfo{person}{Debasis Ganguly}, {and} \bibinfo{person}{Iadh Ounis}.} \bibinfo{year}{2025}\natexlab{}.
\newblock \showarticletitle{One Size Doesn't Fit All: Predicting the Number of Examples for In-Context Learning}. In \bibinfo{booktitle}{\emph{{ECIR} {(1)}}} \emph{(\bibinfo{series}{Lecture Notes in Computer Science}, Vol.~\bibinfo{volume}{15572})}. \bibinfo{publisher}{Springer}, \bibinfo{pages}{67--84}.
\newblock


\bibitem[Dettmers et~al\mbox{.}(2023)]%
        {dettmers2024qlora}
\bibfield{author}{\bibinfo{person}{Tim Dettmers}, \bibinfo{person}{Artidoro Pagnoni}, \bibinfo{person}{Ari Holtzman}, {and} \bibinfo{person}{Luke Zettlemoyer}.} \bibinfo{year}{2023}\natexlab{}.
\newblock \showarticletitle{QLoRA: Efficient Finetuning of Quantized LLMs}. In \bibinfo{booktitle}{\emph{Advances in Neural Information Processing Systems 36: Annual Conference on Neural Information Processing Systems 2023, NeurIPS 2023, New Orleans, LA, USA, December 10 - 16, 2023}}, \bibfield{editor}{\bibinfo{person}{Alice Oh}, \bibinfo{person}{Tristan Naumann}, \bibinfo{person}{Amir Globerson}, \bibinfo{person}{Kate Saenko}, \bibinfo{person}{Moritz Hardt}, {and} \bibinfo{person}{Sergey Levine}} (Eds.).
\newblock
\urldef\tempurl%
\url{http://papers.nips.cc/paper\_files/paper/2023/hash/1feb87871436031bdc0f2beaa62a049b-Abstract-Conference.html}
\showURL{%
\tempurl}


\bibitem[Dubey et~al\mbox{.}(2024)]%
        {dubey2024llama}
\bibfield{author}{\bibinfo{person}{Abhimanyu Dubey}, \bibinfo{person}{Abhinav Jauhri}, \bibinfo{person}{Abhinav Pandey}, \bibinfo{person}{Abhishek Kadian}, \bibinfo{person}{Ahmad Al-Dahle}, \bibinfo{person}{Aiesha Letman}, \bibinfo{person}{Akhil Mathur}, \bibinfo{person}{Alan Schelten}, \bibinfo{person}{Amy Yang}, \bibinfo{person}{Angela Fan}, {et~al\mbox{.}}} \bibinfo{year}{2024}\natexlab{}.
\newblock \showarticletitle{The llama 3 herd of models}.
\newblock \bibinfo{journal}{\emph{ArXiv preprint}}  \bibinfo{volume}{abs/2407.21783} (\bibinfo{year}{2024}).
\newblock
\urldef\tempurl%
\url{https://arxiv.org/abs/2407.21783}
\showURL{%
\tempurl}


\bibitem[Jiang et~al\mbox{.}(2023)]%
        {jiang2023mistral}
\bibfield{author}{\bibinfo{person}{Albert~Q Jiang}, \bibinfo{person}{Alexandre Sablayrolles}, \bibinfo{person}{Arthur Mensch}, \bibinfo{person}{Chris Bamford}, \bibinfo{person}{Devendra~Singh Chaplot}, \bibinfo{person}{Diego de~las Casas}, \bibinfo{person}{Florian Bressand}, \bibinfo{person}{Gianna Lengyel}, \bibinfo{person}{Guillaume Lample}, \bibinfo{person}{Lucile Saulnier}, {et~al\mbox{.}}} \bibinfo{year}{2023}\natexlab{}.
\newblock \showarticletitle{Mistral 7B}.
\newblock \bibinfo{journal}{\emph{ArXiv preprint}}  \bibinfo{volume}{abs/2310.06825} (\bibinfo{year}{2023}).
\newblock
\urldef\tempurl%
\url{https://arxiv.org/abs/2310.06825}
\showURL{%
\tempurl}


\bibitem[Levy et~al\mbox{.}(2023)]%
        {levy2023diversedemonstrationsimproveincontext}
\bibfield{author}{\bibinfo{person}{Itay Levy}, \bibinfo{person}{Ben Bogin}, {and} \bibinfo{person}{Jonathan Berant}.} \bibinfo{year}{2023}\natexlab{}.
\newblock \showarticletitle{Diverse Demonstrations Improve In-context Compositional Generalization}. In \bibinfo{booktitle}{\emph{Proceedings of the 61st Annual Meeting of the Association for Computational Linguistics (Volume 1: Long Papers)}}, \bibfield{editor}{\bibinfo{person}{Anna Rogers}, \bibinfo{person}{Jordan Boyd-Graber}, {and} \bibinfo{person}{Naoaki Okazaki}} (Eds.). \bibinfo{publisher}{Association for Computational Linguistics}, \bibinfo{address}{Toronto, Canada}, \bibinfo{pages}{1401--1422}.
\newblock
\href{https://doi.org/10.18653/v1/2023.acl-long.78}{doi:\nolinkurl{10.18653/v1/2023.acl-long.78}}


\bibitem[Li and Qiu(2023)]%
        {li2023finding}
\bibfield{author}{\bibinfo{person}{Xiaonan Li} {and} \bibinfo{person}{Xipeng Qiu}.} \bibinfo{year}{2023}\natexlab{}.
\newblock \showarticletitle{Finding Support Examples for In-Context Learning}. In \bibinfo{booktitle}{\emph{Findings of the Association for Computational Linguistics: EMNLP 2023}}, \bibfield{editor}{\bibinfo{person}{Houda Bouamor}, \bibinfo{person}{Juan Pino}, {and} \bibinfo{person}{Kalika Bali}} (Eds.). \bibinfo{publisher}{Association for Computational Linguistics}, \bibinfo{address}{Singapore}, \bibinfo{pages}{6219--6235}.
\newblock
\href{https://doi.org/10.18653/v1/2023.findings-emnlp.411}{doi:\nolinkurl{10.18653/v1/2023.findings-emnlp.411}}


\bibitem[Li et~al\mbox{.}(2023)]%
        {li2023textbooks}
\bibfield{author}{\bibinfo{person}{Yuanzhi Li}, \bibinfo{person}{S{\'e}bastien Bubeck}, \bibinfo{person}{Ronen Eldan}, \bibinfo{person}{Allie Del~Giorno}, \bibinfo{person}{Suriya Gunasekar}, {and} \bibinfo{person}{Yin~Tat Lee}.} \bibinfo{year}{2023}\natexlab{}.
\newblock \showarticletitle{Textbooks are all you need ii: phi-1.5 technical report}.
\newblock \bibinfo{journal}{\emph{ArXiv preprint}}  \bibinfo{volume}{abs/2309.05463} (\bibinfo{year}{2023}).
\newblock
\urldef\tempurl%
\url{https://arxiv.org/abs/2309.05463}
\showURL{%
\tempurl}


\bibitem[Ma et~al\mbox{.}(2023)]%
        {ma2023fairness}
\bibfield{author}{\bibinfo{person}{Huan Ma}, \bibinfo{person}{Changqing Zhang}, \bibinfo{person}{Yatao Bian}, \bibinfo{person}{Lemao Liu}, \bibinfo{person}{Zhirui Zhang}, \bibinfo{person}{Peilin Zhao}, \bibinfo{person}{Shu Zhang}, \bibinfo{person}{Huazhu Fu}, \bibinfo{person}{Qinghua Hu}, {and} \bibinfo{person}{Bingzhe Wu}.} \bibinfo{year}{2023}\natexlab{}.
\newblock \showarticletitle{Fairness-guided Few-shot Prompting for Large Language Models}. In \bibinfo{booktitle}{\emph{Advances in Neural Information Processing Systems 36: Annual Conference on Neural Information Processing Systems 2023, NeurIPS 2023, New Orleans, LA, USA, December 10 - 16, 2023}}, \bibfield{editor}{\bibinfo{person}{Alice Oh}, \bibinfo{person}{Tristan Naumann}, \bibinfo{person}{Amir Globerson}, \bibinfo{person}{Kate Saenko}, \bibinfo{person}{Moritz Hardt}, {and} \bibinfo{person}{Sergey Levine}} (Eds.).
\newblock
\urldef\tempurl%
\url{http://papers.nips.cc/paper\_files/paper/2023/hash/8678da90126aa58326b2fc0254b33a8c-Abstract-Conference.html}
\showURL{%
\tempurl}


\bibitem[Maxwell et~al\mbox{.}(2019)]%
        {maxwell2019impact}
\bibfield{author}{\bibinfo{person}{David Maxwell}, \bibinfo{person}{Leif Azzopardi}, {and} \bibinfo{person}{Yashar Moshfeghi}.} \bibinfo{year}{2019}\natexlab{}.
\newblock \showarticletitle{The impact of result diversification on search behaviour and performance}.
\newblock \bibinfo{journal}{\emph{Information Retrieval Journal}} \bibinfo{volume}{22}, \bibinfo{number}{5} (\bibinfo{year}{2019}), \bibinfo{pages}{422--446}.
\newblock


\bibitem[Parry et~al\mbox{.}(2024)]%
        {Parry_2024}
\bibfield{author}{\bibinfo{person}{Andrew Parry}, \bibinfo{person}{Debasis Ganguly}, {and} \bibinfo{person}{Manish Chandra}.} \bibinfo{year}{2024}\natexlab{}.
\newblock \showarticletitle{“In-Context Learning” or: How I learned to stop worrying and love “Applied Information Retrieval”}. In \bibinfo{booktitle}{\emph{Proceedings of the 47th International ACM SIGIR Conference on Research and Development in Information Retrieval}} \emph{(\bibinfo{series}{SIGIR 2024})}. \bibinfo{publisher}{ACM}.
\newblock
\href{https://doi.org/10.1145/3626772.3657842}{doi:\nolinkurl{10.1145/3626772.3657842}}


\bibitem[Pickett et~al\mbox{.}(2024)]%
        {pickett2024better}
\bibfield{author}{\bibinfo{person}{Marc Pickett}, \bibinfo{person}{Jeremy Hartman}, \bibinfo{person}{Ayan~Kumar Bhowmick}, \bibinfo{person}{Raquib-ul Alam}, {and} \bibinfo{person}{Aditya Vempaty}.} \bibinfo{year}{2024}\natexlab{}.
\newblock \showarticletitle{Better RAG using Relevant Information Gain}.
\newblock \bibinfo{journal}{\emph{ArXiv preprint}}  \bibinfo{volume}{abs/2407.12101} (\bibinfo{year}{2024}).
\newblock
\urldef\tempurl%
\url{https://arxiv.org/abs/2407.12101}
\showURL{%
\tempurl}


\bibitem[Reimers and Gurevych(2019)]%
        {reimers2019sentence}
\bibfield{author}{\bibinfo{person}{Nils Reimers} {and} \bibinfo{person}{Iryna Gurevych}.} \bibinfo{year}{2019}\natexlab{}.
\newblock \showarticletitle{Sentence-{BERT}: Sentence Embeddings using {S}iamese {BERT}-Networks}. In \bibinfo{booktitle}{\emph{Proceedings of the 2019 Conference on Empirical Methods in Natural Language Processing and the 9th International Joint Conference on Natural Language Processing (EMNLP-IJCNLP)}}, \bibfield{editor}{\bibinfo{person}{Kentaro Inui}, \bibinfo{person}{Jing Jiang}, \bibinfo{person}{Vincent Ng}, {and} \bibinfo{person}{Xiaojun Wan}} (Eds.). \bibinfo{publisher}{Association for Computational Linguistics}, \bibinfo{address}{Hong Kong, China}, \bibinfo{pages}{3982--3992}.
\newblock
\href{https://doi.org/10.18653/v1/D19-1410}{doi:\nolinkurl{10.18653/v1/D19-1410}}


\bibitem[Roy et~al\mbox{.}(2019)]%
        {DBLP:journals/ipm/RoyGMJ19}
\bibfield{author}{\bibinfo{person}{Dwaipayan Roy}, \bibinfo{person}{Debasis Ganguly}, \bibinfo{person}{Mandar Mitra}, {and} \bibinfo{person}{Gareth J.~F. Jones}.} \bibinfo{year}{2019}\natexlab{}.
\newblock \showarticletitle{Estimating Gaussian mixture models in the local neighbourhood of embedded word vectors for query performance prediction}.
\newblock \bibinfo{journal}{\emph{Inf. Process. Manag.}} \bibinfo{volume}{56}, \bibinfo{number}{3} (\bibinfo{year}{2019}), \bibinfo{pages}{1026--1045}.
\newblock


\bibitem[Rubin et~al\mbox{.}(2022)]%
        {rubin2022learningretrievepromptsincontext}
\bibfield{author}{\bibinfo{person}{Ohad Rubin}, \bibinfo{person}{Jonathan Herzig}, {and} \bibinfo{person}{Jonathan Berant}.} \bibinfo{year}{2022}\natexlab{}.
\newblock \showarticletitle{Learning To Retrieve Prompts for In-Context Learning}. In \bibinfo{booktitle}{\emph{Proceedings of the 2022 Conference of the North American Chapter of the Association for Computational Linguistics: Human Language Technologies}}, \bibfield{editor}{\bibinfo{person}{Marine Carpuat}, \bibinfo{person}{Marie-Catherine de~Marneffe}, {and} \bibinfo{person}{Ivan~Vladimir Meza~Ruiz}} (Eds.). \bibinfo{publisher}{Association for Computational Linguistics}, \bibinfo{address}{Seattle, United States}, \bibinfo{pages}{2655--2671}.
\newblock
\href{https://doi.org/10.18653/v1/2022.naacl-main.191}{doi:\nolinkurl{10.18653/v1/2022.naacl-main.191}}


\bibitem[Santos et~al\mbox{.}(2010)]%
        {santos2010exploiting}
\bibfield{author}{\bibinfo{person}{Rodrygo L.~T. Santos}, \bibinfo{person}{Craig Macdonald}, {and} \bibinfo{person}{Iadh Ounis}.} \bibinfo{year}{2010}\natexlab{}.
\newblock \showarticletitle{Exploiting query reformulations for web search result diversification}. In \bibinfo{booktitle}{\emph{Proceedings of the 19th International Conference on World Wide Web, {WWW} 2010, Raleigh, North Carolina, USA, April 26-30, 2010}}, \bibfield{editor}{\bibinfo{person}{Michael Rappa}, \bibinfo{person}{Paul Jones}, \bibinfo{person}{Juliana Freire}, {and} \bibinfo{person}{Soumen Chakrabarti}} (Eds.). \bibinfo{publisher}{{ACM}}, \bibinfo{pages}{881--890}.
\newblock
\href{https://doi.org/10.1145/1772690.1772780}{doi:\nolinkurl{10.1145/1772690.1772780}}


\bibitem[Sinhababu et~al\mbox{.}(2024)]%
        {sinhababu2024few}
\bibfield{author}{\bibinfo{person}{Nilanjan Sinhababu}, \bibinfo{person}{Andrew Parry}, \bibinfo{person}{Debasis Ganguly}, \bibinfo{person}{Debasis Samanta}, {and} \bibinfo{person}{Pabitra Mitra}.} \bibinfo{year}{2024}\natexlab{}.
\newblock \showarticletitle{Few-shot Prompting for Pairwise Ranking: An Effective Non-Parametric Retrieval Model}.
\newblock \bibinfo{journal}{\emph{ArXiv preprint}}  \bibinfo{volume}{abs/2409.17745} (\bibinfo{year}{2024}).
\newblock
\urldef\tempurl%
\url{https://arxiv.org/abs/2409.17745}
\showURL{%
\tempurl}


\bibitem[Tian et~al\mbox{.}(2025)]%
        {DBLP:conf/ecir/TianGM25}
\bibfield{author}{\bibinfo{person}{Fangzheng Tian}, \bibinfo{person}{Debasis Ganguly}, {and} \bibinfo{person}{Craig Macdonald}.} \bibinfo{year}{2025}\natexlab{}.
\newblock \showarticletitle{Is Relevance Propagated from Retriever to Generator in RAG?}. In \bibinfo{booktitle}{\emph{{ECIR} {(1)}}} \emph{(\bibinfo{series}{Lecture Notes in Computer Science}, Vol.~\bibinfo{volume}{15572})}. \bibinfo{publisher}{Springer}, \bibinfo{pages}{32--48}.
\newblock


\bibitem[Touvron et~al\mbox{.}(2023)]%
        {touvron2023llama}
\bibfield{author}{\bibinfo{person}{Hugo Touvron}, \bibinfo{person}{Louis Martin}, \bibinfo{person}{Kevin Stone}, \bibinfo{person}{Peter Albert}, \bibinfo{person}{Amjad Almahairi}, \bibinfo{person}{Yasmine Babaei}, \bibinfo{person}{Nikolay Bashlykov}, \bibinfo{person}{Soumya Batra}, \bibinfo{person}{Prajjwal Bhargava}, \bibinfo{person}{Shruti Bhosale}, {et~al\mbox{.}}} \bibinfo{year}{2023}\natexlab{}.
\newblock \showarticletitle{Llama 2: Open foundation and fine-tuned chat models}.
\newblock \bibinfo{journal}{\emph{ArXiv preprint}}  \bibinfo{volume}{abs/2307.09288} (\bibinfo{year}{2023}).
\newblock
\urldef\tempurl%
\url{https://arxiv.org/abs/2307.09288}
\showURL{%
\tempurl}


\bibitem[Wang et~al\mbox{.}(2019)]%
        {wang2018glue}
\bibfield{author}{\bibinfo{person}{Alex Wang}, \bibinfo{person}{Amanpreet Singh}, \bibinfo{person}{Julian Michael}, \bibinfo{person}{Felix Hill}, \bibinfo{person}{Omer Levy}, {and} \bibinfo{person}{Samuel~R. Bowman}.} \bibinfo{year}{2019}\natexlab{}.
\newblock \showarticletitle{{GLUE:} {A} Multi-Task Benchmark and Analysis Platform for Natural Language Understanding}. In \bibinfo{booktitle}{\emph{7th International Conference on Learning Representations, {ICLR} 2019, New Orleans, LA, USA, May 6-9, 2019}}. \bibinfo{publisher}{OpenReview.net}.
\newblock
\urldef\tempurl%
\url{https://openreview.net/forum?id=rJ4km2R5t7}
\showURL{%
\tempurl}


\bibitem[Ye et~al\mbox{.}(2023)]%
        {ye2023compositional}
\bibfield{author}{\bibinfo{person}{Jiacheng Ye}, \bibinfo{person}{Zhiyong Wu}, \bibinfo{person}{Jiangtao Feng}, \bibinfo{person}{Tao Yu}, {and} \bibinfo{person}{Lingpeng Kong}.} \bibinfo{year}{2023}\natexlab{}.
\newblock \showarticletitle{Compositional Exemplars for In-context Learning}. In \bibinfo{booktitle}{\emph{International Conference on Machine Learning, {ICML} 2023, 23-29 July 2023, Honolulu, Hawaii, {USA}}} \emph{(\bibinfo{series}{Proceedings of Machine Learning Research}, Vol.~\bibinfo{volume}{202})}, \bibfield{editor}{\bibinfo{person}{Andreas Krause}, \bibinfo{person}{Emma Brunskill}, \bibinfo{person}{Kyunghyun Cho}, \bibinfo{person}{Barbara Engelhardt}, \bibinfo{person}{Sivan Sabato}, {and} \bibinfo{person}{Jonathan Scarlett}} (Eds.). \bibinfo{publisher}{{PMLR}}, \bibinfo{pages}{39818--39833}.
\newblock
\urldef\tempurl%
\url{https://proceedings.mlr.press/v202/ye23c.html}
\showURL{%
\tempurl}


\bibitem[Zhang et~al\mbox{.}(2022)]%
        {zhang2022active}
\bibfield{author}{\bibinfo{person}{Yiming Zhang}, \bibinfo{person}{Shi Feng}, {and} \bibinfo{person}{Chenhao Tan}.} \bibinfo{year}{2022}\natexlab{}.
\newblock \showarticletitle{Active Example Selection for In-Context Learning}. In \bibinfo{booktitle}{\emph{Proceedings of the 2022 Conference on Empirical Methods in Natural Language Processing}}, \bibfield{editor}{\bibinfo{person}{Yoav Goldberg}, \bibinfo{person}{Zornitsa Kozareva}, {and} \bibinfo{person}{Yue Zhang}} (Eds.). \bibinfo{publisher}{Association for Computational Linguistics}, \bibinfo{address}{Abu Dhabi, United Arab Emirates}, \bibinfo{pages}{9134--9148}.
\newblock
\href{https://doi.org/10.18653/v1/2022.emnlp-main.622}{doi:\nolinkurl{10.18653/v1/2022.emnlp-main.622}}


\bibitem[Zhang et~al\mbox{.}(2023)]%
        {zhang2022automatic}
\bibfield{author}{\bibinfo{person}{Zhuosheng Zhang}, \bibinfo{person}{Aston Zhang}, \bibinfo{person}{Mu Li}, {and} \bibinfo{person}{Alex Smola}.} \bibinfo{year}{2023}\natexlab{}.
\newblock \showarticletitle{Automatic Chain of Thought Prompting in Large Language Models}. In \bibinfo{booktitle}{\emph{The Eleventh International Conference on Learning Representations, {ICLR} 2023, Kigali, Rwanda, May 1-5, 2023}}. \bibinfo{publisher}{OpenReview.net}.
\newblock
\urldef\tempurl%
\url{https://openreview.net/pdf?id=5NTt8GFjUHkr}
\showURL{%
\tempurl}


\end{thebibliography}

\appendix

\end{document}